\documentclass[aps,prl,twocolumn,twoside,floatfix,superscriptaddress,preprintnumbers]{revtex4}

\preprint{OUTP-03-18P} 
\preprint{DFPD-03/TH-26}
\preprint{ROMA-1356/03}
\preprint{RM3-TH/03-11}
\usepackage{graphicx}
\newcommand{\bea}{\begin{eqnarray}}
\newcommand{\beq}{\begin{equation}}
\newcommand{\eea}{\end{eqnarray}}
\newcommand{\eeq}{\end{equation}}

\newcommand{\lsim}{\raise0.3ex\hbox{$\;<$\kern-0.75em\raise-1.1ex\hbox{$\sim\;$}}}
\newcommand{\gsim}{\raise0.3ex\hbox{$\;>$\kern-0.75em\raise-1.1ex\hbox{$\sim\;$}}}

\newcommand{\unity}{{\hbox{1\kern-.8mm l}}}

\begin{document}

\title{Grand Unification of Quark and Lepton FCNCs}
\author{M. Ciuchini}
\affiliation{INFN, Sezione di Roma III and Dip. di Fisica, Univ. di Roma Tre,
Via della Vasca Navale 84, I-00146 Rome, Italy.}
\author{A. Masiero}
\affiliation{ Dip. di Fisica `G. Galilei', Univ. di Padova and 
INFN, Sezione di Padova, Via Marzolo 8, I-35131, Padua, Italy.} 
\author{L. Silvestrini}
\affiliation{INFN, Sezione di Roma and Dip. di Fisica, Univ. di Roma
`La Sapienza', P.le A. Moro 2, I-00185 Rome, Italy.}
\author{S. K. Vempati}
\affiliation{ Dip. di Fisica `G. Galilei', Univ. di Padova and 
INFN, Sezione di Padova, Via Marzolo 8, I-35131, Padua, Italy.} 
\author{O. Vives}
\affiliation{Dep. of Physics, U. of Oxford, 1 Keble Road, Oxford, OX1 3NP, UK.}

\begin{abstract}
In the context of Supersymmetric Grand Unified theories with soft
breaking terms arising at the Planck scale, it is generally possible 
to link flavor changing neutral current and CP violating processes occurring 
in the leptonic and hadronic sectors. We study the correlation between flavor 
changing squark and slepton mass insertions in models \`a la 
$SU(5)$. We show that the constraints coming from lepton flavor 
violation exhibit a strong impact on CP-violating $B$ decays.  
\end{abstract}

\maketitle


The phenomenology of low-energy supersymmetric (SUSY) extensions of
the Standard Model (SM) crucially depends on the mechanism responsible
for breaking SUSY itself, namely on the structure of the SUSY soft
breaking terms. It is possible to explore these terms by studying the
production and detection of SUSY particles. In addition, one can shed
light on these terms by analyzing the SUSY effects on rare flavor
changing neutral current (FCNC) and CP violating processes.  Obviously
for this latter approach the flavor structure of the soft breaking
terms is of utmost importance. Theoretically, we can distinguish two
opposite situations. First, we can have SUSY breaking mechanisms which
are completely insensitive to flavor (for instance, pure dilaton
breaking in supergravities, gauge or anomaly mediated SUSY
breaking). Alternatively, SUSY breaking can feel the flavor structure
leading to soft breaking terms which are not flavor universal (this is
what happens in supergravities which have also moduli breaking). In
spite of the original flavor blindness of certain SUSY breaking
mechanisms, we can still have large flavor non-universalities in the
low-energy SUSY breaking sector.  This occurs if the universal soft
terms appear at some energy scale much larger than the electroweak
scale in the presence of new flavor structures (for instance, in 
Grand Unified models or in the presence of new Yukawa couplings as
in the seesaw model). In the running
between these two scales, the soft breaking terms can `pick up'
non-universal contributions: soft scalar masses and scalar trilinear
terms bear a `memory' of the high energy flavor structures through
their RG evolution. 
In conclusion, either that we start from SUSY
breaking mechanisms which are not flavor blind or that we produce
flavor non-universality through the RG scaling, we can envisage the
general situation that soft SUSY breaking terms are flavor
non-universal at the electroweak scale.

To analyze flavor violating constraints at the electroweak scale, the
model independent Mass-Insertion (MI) approximation is advantageous
\cite{kostelesky}. In this method, the experimental limits lead to
upper bounds on the parameters (or combinations of) $\delta_{ij}^f
\equiv \Delta^f_{ij}/m_{\tilde{f}}^2$; where $\Delta^f_{ij}$ is the
flavor-violating off-diagonal entry appearing in the $f = (u,d,l)$
sfermion mass matrices and $m_{\tilde{f}}^2$ is the average sfermion
mass. In addition, the mass-insertions are further sub-divided into
LL/LR/RL/RR types, labelled by the chirality of the corresponding SM
fermions.  Detailed bounds on the individual $\delta$s have been
derived by considering limits from various FCNC processes
\cite{gabbiani}.  As long as one remains within the simple picture of
the Minimal Supersymmetric Standard Model (MSSM), where quarks and
leptons are unrelated, the hadronic and leptonic FCNC processes yield
bounds on corresponding $\delta^q$'s and $\delta^l$'s which are
unrelated.  The situation changes when one embeds the MSSM within a
Grand Unified Theory (GUT).

In this letter, we study the correlation between the $\delta^q$'s and
$\delta^l$'s in SUSY GUTs and show that lepton flavor violating
processes (as $\mu \to e\,\gamma $ or $ \tau \to \mu\, \gamma$) can
severely constrain the observability of SUSY contributions to hadronic
FCNC processes and vice-versa.

In a SUSY GUT, quarks and leptons sit in the same multiplet.  As long
as the scale associated with the transmission of SUSY breaking to the
visible sector is larger than the GUT breaking scale, the quark-lepton
unification seeps also into the SUSY breaking soft sector leading to
squark-slepton mass squared unification. The exact relations between
the mass matrices depend on the choice of the GUT gauge group. As a
consequence, the flavor off-diagonal entries $\Delta_{ij}^f$ in these
mass matrices are also unified at the GUT scale. For instance, in
$SU(5)$ $(\Delta^d_{ij})_{RR}$ and $(\Delta^l_{ij})_{LL}$ are equal;
in $SO(10)$ all $\Delta_{ij}$ are equal at $M_{GUT}$ implying strong
correlations within FCNCs at that scale.  What is going to happen of
such strong correlations when we reach the electroweak scale? In a
simple situation of ``Big Desert'', \textit{i.e.}, no new particles or
interactions between $M_{GUT}$ and $M_W$, these relations are left
essentially unaffected by the RGE effects and thus the relations
between $\Delta_{ij}$ can be cast in terms of $\delta_{ij}$ at the
weak scale. The situation changes if some new particles with
intermediate masses interact differently with quarks and leptons, for
instance right-handed neutrinos in a see-saw mechanism. In this
particular case, the leptonic $\Delta_{ij}$ receive new RGE
contributions \cite{fbam}, breaking the exact hadron-lepton
correlation at the electroweak scale. However, as we show below, it is
still possible to extract useful information from these GUT
correlations.

The connection between quark and lepton $\delta$ parameters can have
significant implications on flavor phenomenology. Indeed, using these
relations, a quark $\delta$ parameter can be probed in a leptonic
process or vice versa. In this way, it is possible that constraints in
one sector are converted to the other sector where previously only
weaker or perhaps even no bounds existed. For example, we find that
using the leptonic radiative decay process $l_j \to l_i\, \gamma$, it
is possible to probe some squark $\delta$ parameters to a higher
precision, unhindered by the hadronic uncertainties.  On the other
hand, some sleptonic $\delta$ parameters, which are unconstrained, can
receive bounds from the quark sector.

To be specific, we concentrate on the SUSY $SU(5)$ framework and
derive all the relations between squark and sleptonic mass
insertions. We then study the impact of the limit from $\tau \rightarrow
\mu\, \gamma$ on the $b \to s $ transition observables, such as
$A_{CP} (B \to \phi K_s)$.

The soft terms are assumed to be generated at some scale above
$M_{GUT}$.  Note that even assuming complete universality of the soft
breaking terms at $M_{Planck}$, as in mSUGRA, the RG effects to
$M_{GUT}$ will induce flavor off-diagonal entries at the GUT scale
\cite{barbieri}.  Hence we assume generic flavor violating entries to
be present in the sfermion matrices at the GUT scale.  The part of the
superpotential relevant for quarks and charged lepton masses
can be
written as
\beq
\label{su5super}
W_{SU(5)} = h^u_{ij}~ T_i~ T_j~ H + h^d_{ij} T_i~ \bar{F}_j~ \bar{H}
+ \mu~ H~ \bar{H} ,
\eeq 
where we have used the standard notation with $T$ transforming as 
$10$  and $\bar{F}$  as $\bar{5}$ under $SU(5)$.
The corresponding $SU(5)$ invariant soft potential has now the form : 

\bea
\label{su5soft}
&&- {\cal L}_{soft} = m_{T_{ij}}^2 \tilde{T}_i^\dagger \tilde{T}_j + 
m_{\bar{F}}^2 \tilde{\bar{F}}_i^\dagger \tilde{\bar{F}}_j +m^2_{H} 
+ m^2_{\bar{H}} \nonumber \\ 
&&~~ + A^u_{ij}~ T_i~ T_j~ H + A^d_{ij}~ T_i~\bar{F}_j~ \bar{H}
+ B \mu ~ H~ \bar{H}, 
\eea 
where the same symbols are used for the scalar components of the superfields.
Rewriting the above in terms of the Standard Model representations we have 
\bea 
\label{smsoft}
- {\cal L}_{soft} = m_{Q_{ij}}^2 \tilde{Q}_i^\dagger \tilde{Q}_j 
+ m_{u^c_{ij}}^2 \tilde{u^c}_i^\star \tilde{u^c}_j + m^2_{e^c_{ij}} 
\tilde{e^c}_i^\star \tilde{e^c}_j ~~~~~\nonumber \\
 + m^2_{d^c_{ij}} \tilde{d^c}^\star_i 
\tilde{d^c}_j + m_{L_{ij}}^2 \tilde{L}_i^\dagger \tilde{L}_j + 
m^2_{H_1} H^\dagger_1 H_1 
+  m^2_{H_2} H_2^\dagger H_2  \nonumber \\
 + A^u_{ij}~
\tilde{Q}_i \tilde{u^c}_j H_2 + A^d_{ij}~
\tilde{Q}_i \tilde{d^c}_j H_1 + A^e_{ij}~
\tilde{L}_i \tilde{e^c}_j H_1 + \ldots ~
\eea
where 
\bea 
\label{matrel1}
m^2_{Q} = m^2_{\tilde{e^c}} = m^2_{\tilde{u^c}} = m^2_{T} \\
\label{matrel2}
m^2_{\tilde{d^c}} = m^2_{L} = m^2_{\bar{F}} \\
\label{trirel}
A^e_{ij} = A^d_{ji}\, .
\eea 
Eqs.~(\ref{matrel1}, \ref{matrel2}, \ref{trirel}) are matrices in flavor 
space.
These 
equations lead to relations within the slepton and squark flavor violating  
off-diagonal entries $\Delta_{ij}$. These are:  
\bea
\label{cdeltas1}
(\Delta^u_{ij})_{LL} = (\Delta^u_{ij})_{RR} = (\Delta^d_{ij})_{LL} = 
(\Delta^l_{ij})_{RR} \\
\label{cdeltas3}
(\Delta^d_{ij})_{RR} = (\Delta^l_{ij})_{LL} \\
\label{cdeltas4}
(\Delta^d_{ij})_{LR} = (\Delta^l_{ji})_{LR} = (\Delta^l_{ij})_{RL}^\star
\eea 
The relations between $\Delta_{LL}$'s and $\Delta_{RR}$'s descend
from the fermionic representations under $SU(5)$, whilst the 
$\Delta_{LR}$'s and $\Delta_{RL}$'s relations also depend on the
Higgs responsible for fermion masses. 

The relations are exact at $M_{GUT}$; however, after $SU(5)$ breaking,
quarks and leptons suffer different renormalization effects and thus,
they are altered at $M_W$.  It is easy to see from the RG equations
that off-diagonal elements in the squark mass matrices in
eqs.~(\ref{cdeltas1} - \ref{cdeltas3}) are approximately not
renormalized due to the smallness of CKM mixing angles and that the
sleptonic entries in them are left unchanged (in the absence of
right-handed neutrinos). On the other hand, eq.~(\ref{cdeltas4})
receives corrections due to the different nature of the RG scaling of
the $LR$ term (A-parameter). This correction can be roughly
approximated as proportional to the corresponding fermion masses.
Taking this into consideration, we can now rewrite the
eqs.~(\ref{cdeltas1} -- \ref{cdeltas4}) at the weak scale as : \bea
\label{deltas1}
(\delta^d_{ij})_{RR} &\approx& {m_{L}^2 \over m_{d^c}^2} (\delta^l_{ij})_{LL}, \\
\label{deltas2}
(\delta^{u,d}_{ij})_{LL}& \approx& {m_{e^c}^2 \over m_{Q}^2} (\delta^l_{ij})_{RR}, \\
\label{deltas3}
(\delta^u_{ij})_{RR} &\approx& {m_{e^c}^2 \over m_{u^c}^2} (\delta^l_{ij})_{RR},\\
\label{deltas4}
(\delta^d_{ij})_{LR}&\approx& {m_{L_{avg}}^2 \over m_{Q_{avg}}^2} ~
{m_b \over m_\tau} ~
(\delta^l_{ij})_{RL}^\star , 
\eea 
where $m_{L_{avg}}^2$ ($m^2_{Q_{avg}}$ ) are given by the geometric average
of left- and right-handed slepton (down-squark) masses
$\sqrt{ m^2_L ~m^2_{e^c}}$ ~~$ \left( \sqrt{m^2_Q ~m^2_{d^c}}~ \right)$,
all of them defined at the weak scale.

To account for non-zero neutrino masses, the seesaw mechanism can be
implemented within this model by adding singlet right-handed
neutrinos. In their presence, additional couplings occur in
eqs.~(\ref{su5super} - \ref{cdeltas4}) at the high scale,
corresponding to the Dirac and Majorana masses which affect the RG
evolution of slepton matrices.  To understand the effect of these new
couplings, one can envisage two scenarios \cite{so10}: (a) small
couplings and/or small mixing in the neutrino Dirac Yukawa matrix, (b)
large couplings and large mixing in the neutrino sector. In case (a),
the effect on slepton mass matrices due neutrino Dirac Yukawa
couplings is very small and the above relations eqs.~(\ref{deltas1}) -
(\ref{deltas4}) still hold. In case (b), however, large RG effects can
significantly modify the slepton doublet flavor structure \cite{fbam}
while keeping the squark sector and right handed charged slepton
matrices essentially unmodified, thus breaking the GUT symmetric
relations. Even in this case, barring accidental cancellations among
the mass insertions already present at $M_{GUT}$ and the radiatively
generated mass insertions between $M_{GUT}$ and $M_{\nu_R}$, there
exists an upper bound on the down quark $\delta$ parameters of the
form:
\begin{equation}
\label{ineqdeltas}
|(\delta^d_{ij})_{RR}| ~~~\leq~~~ {m_{L}^2 \over m_{d^c}^2}
|(\delta^l_{ij})_{LL}|
\end{equation}
while Eqs.~(\ref{deltas2} -- \ref{deltas4}) remain still
valid in this case.

\begingroup
\begin{table}
\begin{ruledtabular}
\begin{tabular}{ccccc}
Transitions (L) && Transitions (D) && Transitions (U) \\[0.2pt]
\hline
$(\delta^l)_{RR}$ && $(\delta^d)_{LL}$ && $(\delta^u)_{LL,RR}$ \\[0.2pt]
\hline
$\mu \to e $ &$\leftrightarrow$&  $s \to d $ & $\leftrightarrow$ &
$ c \to u$  \\
$\tau \to e $ &$\leftrightarrow$& $b \to d $ &$\leftrightarrow$&
$ t \to u$ \\
$\tau \to \mu $ &$\leftrightarrow$& $b \to s $   &$\leftrightarrow$&
$ t \to c$\\
\hline
$(\delta^l)_{LL/LR/RL}$ && $(\delta^d)_{RR/RL/LR}$ && -- \\[0.2pt]
\hline
$\mu \to e $ &$\leftrightarrow$&  $s \to d $ & -- & --  \\
$\tau \to e $ &$\leftrightarrow$& $b \to d $ & -- & --\\
$\tau \to \mu $ &$\leftrightarrow$& $b \to s $ & -- & --
\end{tabular}
\end{ruledtabular}
\caption{Links among various transitions between up-type, down-type quarks
and charged leptons. The corresponding $\delta$'s are shown.}
\label{tb1}
\end{table}
\endgroup

The relations (\ref{deltas2} - \ref{ineqdeltas}) predict links
between lepton and quark flavor changing transitions at the weak
scale. We summarize these links in Table \ref{tb1}.
For example, we see that $\mu \rightarrow e\, \gamma$ can be related to
$K^0-\bar{K}^0$ mixing and to $D^0 -\bar{D}^0$ mixing. Similarly, one
can expect correlations between $\tau \rightarrow e\, \gamma$ and $B_d
- \bar{B}_d$ mixing as well as between  $\tau \rightarrow \mu\, \gamma$
and  $b \to s$ transitions such as $B \to \phi K_s$.

To demonstrate the impact of these relations, let us assume that all
the flavor diagonal sfermion masses are approximately universal at
the GUT scale with $m_T^2 =m_{\tilde F}^2 = m_H^2 = m_{\tilde H}^2 =
m_0^2$ with flavor off-diagonal entries
\begin{equation}
m_{\tilde{f}}^2 = m_0^2 \unity + \Delta^f_{ij}\,, \quad \mathrm{with}
\quad |\Delta^f_{ij}| \leq m_0^2. 
\end{equation}
$\Delta^f_{ij}$ can be present either through the running from the Planck scale
to the GUT scale \cite{barbieri} or through some flavor non-universality
originally present \cite{stringrefs}.  All gaugino masses are unified to 
$M_{1/2}$ at $M_{GUT}$. As mentioned above, the $\Delta$ parameters do not 
receive any significant corrections, whereas the diagonal entries are 
significantly modified (see for instance Tables I and IV in \cite{wien}).
  For a given set of initial conditions ($M_{1/2}$,
$m_0^2$, $A_0$, $\Delta_{ij}$, $\tan \beta$) we obtain the full
spectrum at $M_W$ with the requirement of radiative symmetry breaking.
We then apply limits from direct searches on SUSY particles.
Finally, we calculate the contributions of different $\delta_{23}$
parameters to both leptonic process and hadronic processes, 
considering the region in the $(m_0, M_{1/2})$
plane corresponding to a relatively light sparticle spectrum, 
with squark masses of roughly 350--550 GeV and
slepton masses of about 150--400 GeV.

\begin{figure}
\includegraphics[width=8.6cm]{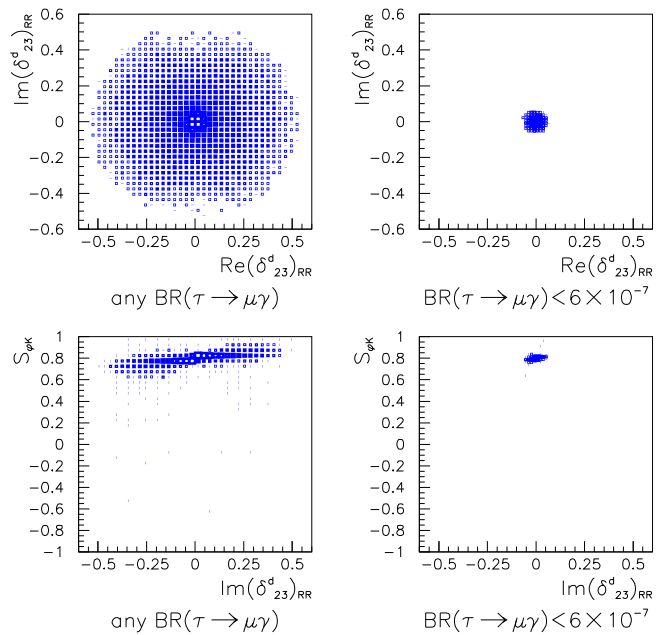}
\caption{
Allowed regions in the 
Re$(\delta^d_{23})_{RR}$--Im$(\delta^d_{23})_{RR}$ plane (top) and
in the $S_{K\phi}$--Im$(\delta^d_{23})_{RR}$ plane (bottom). Constraints
from $B \to X_s \gamma$, $BR(B \to X_s \ell^+ \ell^-)$, and 
the lower bound on $\Delta M_s$ have been used.  }
\label{fig:combi}
\end{figure}

The $b \to s$ transitions have received recently much interest as it
has been shown that the discrepancy with SM expectations in the
measurements of $A_{CP}(B \to \phi K_s)$ can be attributed to the
presence of large neutrino mixing within $SO(10)$ models
\cite{amhm}. Subsequently, a detailed analysis has been
presented \cite{silvest,murayama} within the context of the MSSM. It
has been shown that, for squark and gluino masses around $350$ GeV,
the presence of a $~\mathcal{O}(1)$
$(\delta^d_{23})_{LL,RR}$ could lead to significant discrepancies from
the SM expectations. Similar statements hold for a $~
\mathcal{O}(10^{-2})$ LR or RL MI. In the present work, we study the
impact of LFV bounds on these $\delta$ parameters and subsequently the
effect on B-physics observables. In Table \ref{tb3}, we present upper bounds
on $\delta^d_{23}$ within the above mass ranges for three values of
the limits on Br($\tau \to \mu\, \gamma$). There are no bounds on
$(\delta^d_{23})_{LL}$ because, as is well known \cite{hisano},
large values of $(\delta^l_{ij})_{RR}$ are still allowed due to the
cancellations of bino and higgsino contributions in the decay
amplitude.

At present, the constraints coming from $B$ physics are stronger than
those obtained for the lepton sector in the cases of
$(\delta^d_{23})_{LL,LR,RL}$. Therefore no impact on $B$ phenomenology
is expected even if the present bound on $BR(\tau \to \mu \, \gamma)$
were pushed down to $1 \times 10^{-7}$.  On the contrary, the bound on
$(\delta^d_{23})_{RR}$ induced by $BR(\tau \to \mu \, \gamma)$ is
already at present much stronger than the bounds from hadronic
processes, reducing considerably the room left for SUSY effects in $B$
decays. To illustrate this point in detail, we repeat the analysis of
ref.~\cite{silvest} including the bounds coming from lepton processes.
We therefore compute at the NLO branching ratios and CP asymmetries
for $B \to X_s \gamma$ and $B \to \phi K_s$, $BR(B \to X_s \ell^+
\ell^-)$ and $\Delta M_s$ (see ref.~\cite{silvest} for details). In
the first row of Fig.~\ref{fig:combi}, we plot the probability density in
the Re$(\delta^d_{23})_{RR}$--Im$(\delta^d_{23})_{RR}$ plane for
different upper bounds on $BR(\tau \to \mu \, \gamma)$. Note that
making use of eq.~(\ref{deltas1}) with $|(\delta^l_{23})_{LL}|~ <~ 1
$, implies $\vert (\delta^d_{23})_{RR} \vert \lsim 0.5$ as the ratio
$(m_{L}^2/m_{d^c}^2)$ varies roughly between $(0.2 - 0.5)$ at the weak
scale, for the chosen high scale boundary conditions.  The effect on
$(\delta^d_{23})_{RR}$ of the upper bound on $BR(\tau \to \mu \,
\gamma)$ is dramatic already with the present experimental value.
Correspondingly, as can be seen from the second row of
Fig.~\ref{fig:combi}, the possibility of large deviations from the SM in
the coefficient $S_{\phi K}$ of the sine term in the time-dependent
$A_{CP}(B \to \phi K_s)$ is excluded in the RR case.  Hence, we
conclude that in SUSY GUTs the most likely possibility to strongly
depart from the SM expectations for $S_{\phi K}$ relies on a sizeable
contribution from $(\delta^d_{23})_{LL}$ or $(\delta^d_{23})_{LR,RL}$
as long as they are small enough to be within the severe limits
imposed by $BR(B \to X_s \gamma)$ \cite{silvest}.

\begin{table}
\begin{ruledtabular}
\begin{tabular}{cccc}
Type & $< ~6\cdot 10^{-7}$ & $<~ 1 \cdot 10^{-7}$ & $<~ 1 \cdot 10^{-8}$\\[0.2pt]
\hline
LL & - & - & -   \\
RR & 0.070 & 0.030 & 0.010 \\
RL & 0.080 & 0.035 & 0.010\\
LR & 0.080 & 0.035 & 0.010
\end{tabular}
\end{ruledtabular}
\caption{Bounds on $(\delta^d_{23})$ from $BR(\tau \to \mu \, \gamma)$ for three
different values of the branching ratios for tan $\beta$ = 10.}
\label{tb3}
\end{table}

Exploiting the Grand Unified structure of the theory, we can obtain
similar bounds on other $\delta^d_{ij}$ parameters. For example,
considering the first two generations, the bound on $\delta^d_{12}$
from $BR(\mu \to e\, \gamma)$ can in many cases compete with the bound
from $\Delta m_K$ \cite{gabbiani}.  Similar comparisons can be made
for the $\delta^d_{13}$ from limits on $BR(\tau \to e\, \gamma)$ and
$B_d^0 - \bar{B}_d^0$ mixing. A detailed analysis of $\delta^d_{12}$
and $\delta^d_{13}$ will be presented elsewhere \cite{ournext}.

In summary, Supersymmetric Grand Unification predicts links
 between various leptonic and hadronic FCNC
observables. Though such links have recently been observed 
in the literature within the context of the see-saw
mechanism \cite{othercorrelations}, to our knowledge, it 
is the first time that relations between hadronic and
leptonic mass insertions have been presented in a general
 manner. Though such relations can be constructed for
any GUT group, we have concentrated on $SU(5)$ and quantitatively
 studied the implications for transitions
between the second and third generations. We have shown that 
the present limit on $BR(\tau \to \mu\, \gamma)$ is
sufficient to significantly constrain the observability of 
supersymmetry in CP violating B-decays.

\noindent
\textit{Acknowledgements:}
We acknowledge support from the RTN European project HPRN-CT-2000-0148.
O.V. acknowledges partial support from the Spanish  MCYT FPA2002-00612
and thanks J. March-Russell and G.G. Ross for helpful discussions.

\end{document}